\documentstyle[epsf,twocolumn]{esa_sp_latex}

\def\Ec{E_{\rm c}}
\def\a218{\alpha_{2-18}}

\def\lobs{l_{\rm obs}}
\def\ldiss{l_{\rm diss}}

\def\Ldiss{L_{\rm diss}}
\def\Te{T_{\rm e}}

\def\Lobs{L_{\rm obs}}

\def\alphint{\alpha_{\rm int}}
\def\aint218{\alpha_{{\rm int},2-18}}

\def\Ldiss{L_{\rm diss}}
\def\Lh{L_h}
\def\Ls{L_s}

\def\Tbb{T_{\rm bb}}
\newcommand{\bez}{\begin{eqnarray*}}
\newcommand{\eez}{\end{eqnarray*}}
\newcommand{\be}{\begin{equation}}
\newcommand{\ee}{\end{equation}}

\newcommand{\beq}{\begin{eqnarray}}
\newcommand{\eeq}{\end{eqnarray}}
\newcommand{\bc}{\begin{center}}
\newcommand{\ec}{\end{center}}

\def\ergs{\;{\rm erg}}

\def\cm{\;{\rm cm}}

\def\sec{\;{\rm s}}

\def\taut{\tau_{\rm T}}

\def\sigmat{\sigma_{\rm T}}

\newbox\grsign \setbox\grsign=\hbox{$>$} \newdimen\grdimen \grdimen=\ht\grsign
\newbox\simlessbox \newbox\simgreatbox \newbox\simpropbox
\setbox\simgreatbox=\hbox{\raise.5ex\hbox{$>$}\llap
     {\lower.5ex\hbox{$\sim$}}}\ht1=\grdimen\dp1=0pt
\setbox\simlessbox=\hbox{\raise.5ex\hbox{$<$}\llap
     {\lower.5ex\hbox{$\sim$}}}\ht2=\grdimen\dp2=0pt
\setbox\simpropbox=\hbox{\raise.5ex\hbox{$\propto$}\llap
     {\lower.5ex\hbox{$\sim$}}}\ht2=\grdimen\dp2=0pt

\begin{document}
%

\parindent 0pt
\parskip 10pt plus 1pt minus 1pt
\hoffset=-1.5truecm
\topmargin=-1.0cm
\textwidth 17.1truecm \columnsep 1truecm \columnseprule 0pt 

\topmargin 1cm

\title{\bf TWO-PHASE PAIR CORONA MODEL FOR AGN: 
PHYSICAL MODELLING AND DIAGNOSTICS} 

\author{{\bf Juri~Poutanen$^{1,2}$, 
Roland~Svensson$^1$, 
and Boris~Stern$^{1,3}$} \vspace{2mm} \\
$^1$Stockholm Observatory, Saltsj\"obaden, Sweden\\[1.5mm]
$^2$Uppsala Astronomical Observatory, Uppsala, Sweden\\[1.5mm]
$^3$Institute for Nuclear Research, Moscow, Russia}  

\maketitle

\begin{abstract}
The predictions of the two-phase accretion disc-corona models 
for active galactic nuclei are compared with observations. 
We discuss the possibility to 
use X-ray spectral slopes, equivalent widths of the iron 
line, and the observed flux-spectral index correlation as  diagnostics of the 
X-$\gamma$-ray source compactness and geometry as well as of the 
cold disc temperature. As an example of the application of the 
modelling tools, we use XSPEC to fit 
the broad-band data of Seyfert 1 galaxy, IC4329A,  with 
a theoretical spectrum from a hemisphere-corona.  \vspace{5pt} \\


\end{abstract}

\section{INTRODUCTION}

Observations of Seyfert 1 galaxies by {\em Ginga} and earlier by 
{\em HEAO 1} showed that their X-ray 
spectra consist of at least two distinct components: an intrinsic 
power-law with energy index, $\a218=$0.95$\pm$0.15, in 2 - 18 keV range 
and a Compton reflection bump (\cite{mu93}, \cite{np94}, \cite{w95}). 
Recent {\em ASCA} observations also show these features 
(\cite{n96}). 
{\em OSSE} observations show a much steeper spectrum with $\alpha\approx$ 1.5 
(\cite{joh96}), and {\em COMPTEL} has not detected Seyfert galaxies at all 
(\cite{mai95}). 
Averaged broad-band spectrum for a sample of Seyfert 1 galaxies 
which was obtained using non-simultaneous {\em Ginga} and OSSE data 
shows a high-energy cutoff of the intrinsic power-law at 
$\Ec=$600$^{+800}_{-300}$ keV (\cite{z95}). 
A similar cutoff energy is obtained by \cite*{gon96} using the 
data from {\em EXOSAT} and OSSE. 
{\em ROSAT, Ginga} and OSSE data of the Seyfert 1 galaxy, IC4329A, 
also require a cutoff in the intrinsic spectrum (\cite{ma95}). 
The brightest Seyfert  galaxy in $\gamma$-rays, NGC 4151, shows 
a cutoff at $\sim$ 150 keV, but is probably a freak object requiring 
special considerations (\cite{p96}, but see \cite{zdz97}). 

Non-thermal pair models intensively studied 
in the mid 80s (\cite{sve86}, for review see \cite{sve94})
predicted the spectral index of the intrinsic component 
$\alphint\approx 0.9-1.0$ and a prominent pair annihilation line. 
Absence of the annihilation line in the observed spectra rules out 
pure non-thermal models. Attention is now focused on thermal models.
A definite answer on the question about the relative fraction of the 
thermal and non-thermal plasma cannot be obtained without 
high quality data above 200 keV. 

\begin{figure}[h]
  \begin{center}
    \leavevmode
\epsfysize=5.0cm  
\epsfbox[94 380 478 700]{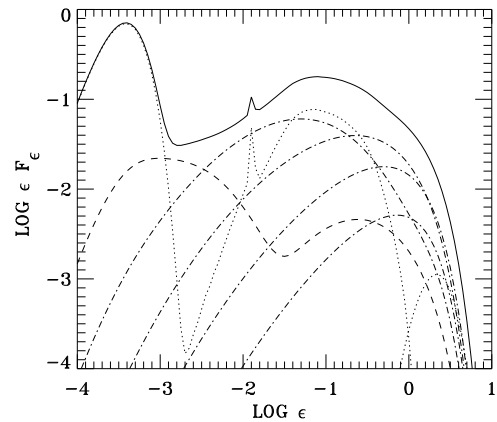}
  \end{center}
  \caption{\em An example of the deconvolution of the emergent spectrum 
from a disc-corona system 
into different  scattering orders. Solid curve represents the total 
emergent spectrum. Dotted curves give the contribution of the 
unscattered radiation  which consists of three components: black-body 
radiation with a maximum at 
$\epsilon\equiv h\nu/m_ec^2\approx$3$\cdot$10$^{-4}$, 
a Compton reflection component with a maximum at $\epsilon\approx$0.1, 
and a broad annihilation line at $\epsilon\approx$2. 
Dashed curve gives the contribution of the single scattered radiation
(notice that the  broad bump with a maximum at $\epsilon\approx$0.2 is
the first scattering order of the Compton reflection component). 
All other scattering orders are represented by dash-dotted curves. } 
 \label{fig:decon} 
\end{figure}

There is a consensus that the  X/$\gamma$-ray spectrum of Seyferts
is produced by Comptonization in hot plasmas of soft radiation from 
the UV source. The  exact geometry of both phases is unknown. 
Presently, the most commonly used scenario 
is the two-phase disc-corona model
(e.g., \cite{hm91}, 1993) in which a hot X-ray emitting corona
is located above the cold UV-emitting disc of the canonical
black hole model for AGNs. The power-law X-ray spectrum
is generated by thermal Comptonization of the soft UV-radiation.
About half of the X-rays enters and is reprocessed by the cold disc,
emerging mostly as black body disc radiation in the UV (some fraction 
is reflected producing a Compton reflection component). 
\cite*{hm91} emphasized the coupling between the
two phases due to the reprocessing, as the soft disc photons
influence the cooling of the corona. They showed that nearly all
power must be dissipated in the corona in order to have
$\alphint \sim$ 0.9-1.
 A consequence of this is that the soft disc
luminosity, $\Ls$, is of the same order as the hard X-ray luminosity,
$\Lh$.

\begin{figure}[htbp] 
  \begin{center}
    \leavevmode
\epsfxsize=6.0cm  
\epsfbox[164 380 418 720]{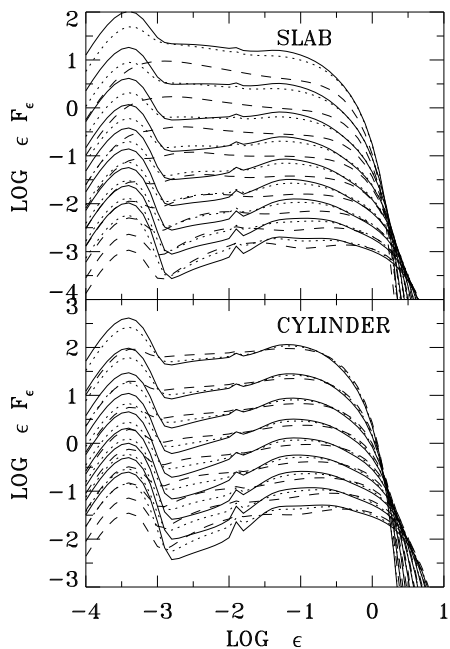}
  \end{center}
  \caption{\em Upper panel: spectra emerging in different directions
from a slab-corona. Solid, dotted, and dashed  curves represent 
the flux at viewing cosine angle $\mu$=0.9, 0.5 and 0.1, respectively. 
The temperature of the black-body radiation from the accretion disc 
is fixed at $\Tbb$=50 eV. The electron temperature of the corona 
increases from the top of the figure to the bottom:  
$\Theta\equiv k\Te/mc^2$= 0.16, 0.2, 0.25, 0.32, 0.4, 0.5, 0.63, 0.79, 1.0. 
Notice the increase of the iron line equivalent width with increasing 
temperature due to anisotropic scattering effects (see, e.g., Haardt $\&$ 
Maraschi 1993, Stern et al. 1995b). 
The iron line profile has a triangular shape 
due to the energy resolution employed here. Lower panel: 
spectra from a cylinder-corona with a height-to-radius ratio 2. } 
 \label{fig:slab} 
\end{figure}

Observations show that $\Ls$ may be several times larger than
$\Lh$, in contradiction to the prediction of the uniform two-phase
disc-corona model. This led \cite*{hmg94} 
to propose a patchy disc-corona model, where the corona consists
of several localized active regions on the surface of the disc.
Internal disc dissipation results in UV-radiation that leaves
the disc without entering the active regions, thus enhancing
the observed $\Ls$ relative to $\Lh$. 

The first exact radiative transfer/Comptonization  calculations 
in a disc-corona system
accounting for energy and pair balance  as well as reprocessing 
by the cold disc (including angular  anisotropy) were reported in \cite{s95b}. 
Two methods  were used.  
The first method  is based on the Non-Linear Monte-Carlo 
method by Stern (see detailed description in \cite{s95a}). The second method 
is the iterative scattering method (ISM) where the radiative transfer is 
exactly solved  for each scattering order separately (\cite{ps96}). 
The results of both codes are in excellent agreement.
In this paper we report the results using the second method.  
A typical spectrum emerging from the disc-corona system and 
its different components is shown in Figure~\ref{fig:decon}.

\section{SPECTRA FROM THE DISC-CORONA SYSTEM} 
\label{sec:spectr} 

The ISM code is a 1D code but it can also treat quasi-2D radiative transfer
in cylinders and hemispheres. The full Compton scattering
matrix is used allowing us to treat  polarized radiative transfer
in thermal relativistic plasmas. Fully angular dependent, polarized
Compton reflection is implemented using a Green's matrix (\cite{pns96}). 
For a given geometry, energy balance gives a unique 
combination of electron temperature $\Theta=k\Te/m_ec^2$ and 
Thomson optical depth, $\taut=n_e\sigmat h$.  
Solving the pair balance for a given ($\Theta$, $\taut$)
provides  a unique dissipation compactness, $\ldiss$ (\cite{gh94}).   
Here, the local dissipation compactness,
$\ldiss \equiv ( \Ldiss /h ) (\sigmat /  m_e c^3)$,
characterizes the dissipation with $\Ldiss$ being the power providing uniform
heating in a cubic volume of size $h$ in the case of a slab of height $h$,
or in the whole
volume in the case of an active region (cylinder or hemisphere) of height $h$. 
The relation between $\Theta$, $\taut$, and $\ldiss$ for different geometries 
is given in \cite*{s95b} (see also \cite{sve96}).

Figure~\ref{fig:slab} shows  
spectra from slab-corona and cylinder-corona at 
different $\Theta$. 
As the temperature increases the first  order scattered photons are 
predominantly scattered back into the disc, causing a deficiency of the 
first order photons in the observed spectrum. This anisotropic effect gives
rise to an anisotropy break in the spectra (\cite{s95b}).

\section{DIAGNOSTICS USING SPECTRAL SLOPES AND IRON LINE EQUIVALENT WIDTHS} 
\label{sec:diag}

\begin{figure}[htbp] 
  \begin{center}
    \leavevmode
\epsfxsize=6.0cm
\epsfbox[134 380 418 720]{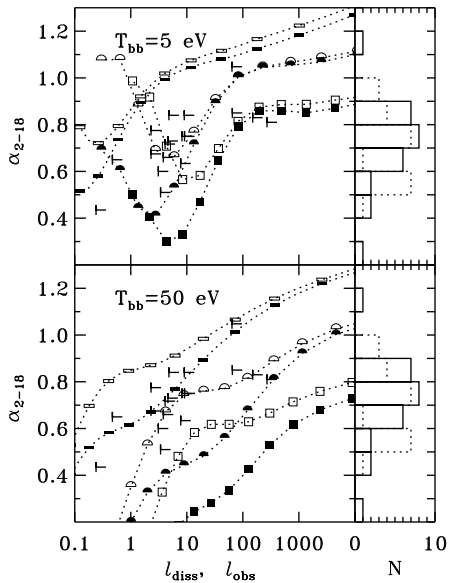}
  \end{center}
  \caption{\em  Overall power-law index, $\a218$, least-square fitted 
to the model spectra in the 2 - 18 keV energy range vs. dissipation 
compactness, $\ldiss\equiv ( \Ldiss /h ) (\sigmat /  m_e c^3)$. 
Rectangles, hemispheres, and 
squares represent results for the slab-corona, hemisphere-corona, and
cylinder-corona (with height-to-radius ratio 2), respectively. 
Filled symbols correspond 
to the spectra in almost face-on direction ($\mu$=0.9), open 
symbols represent results for inclination 60$^{\rm o}$ ($\mu$=0.5).  
Results for $\Tbb$=5 eV and $\Tbb$=50 eV are presented in upper and 
lower panels, respectively. Lower compactness limits represent Seyfert galaxies 
that have known estimates of their X-ray variability, $\Delta t$, and thus 
lower limits of their compactnesses, 
$\lobs\equiv (\Lobs/c\Delta t)(\sigmat/m_ec^3)$, see Done $\&$ Fabian 1989. 
The right panels shows the observed
distributions of power-law indices of  Seyfert galaxies. Solid histogram: 
 Ginga data from Nandra $\&$ Pounds (1994).  Dotted histogram: 
 ASCA data from Nandra et al. (1996, Table~2). } 
 \label{fig:al218} 
\end{figure}

In order to compare the predictions of the two-phase corona model
with observations, we determine 
 the least square overall spectral slope, $\a218$,
for theoretical model spectra and display them  
in Figure~\ref{fig:al218} as a function of the dissipation compactness,
$\ldiss$, for different geometries.
The right panels show the observed distributions of the spectral indices 
for {\em Ginga} spectra 
of 27 Seyfert galaxies (\cite{np94}) and 
for {\em ASCA} spectra (2 - 10 keV range) of 15 Seyfert 1s (\cite{n96}). 
One sees that the observations are more consistent with active surface regions
(such as hemispheres or cylinders)
than with slabs.  Active regions produce spectra covering
the {\em observed ranges} of spectral indices and 
cutoff energies for the {\em observed range} of compactnesses.

For small temperature of the black-body, $\Tbb$,  anisotropic  
effects significantly change the  spectral index at compactnesses 
smaller than $\sim$ 100. For $\Tbb$=50 eV, the anisotropy break 
in the 2 - 18 keV range appears all the way up to $\ldiss\approx$ 1000.
For such a high temperature of the soft radiation, the first scattering 
order extends up to the {\em Ginga} range already for $\Te\approx$100 keV.

\begin{figure}[htbp] 
  \begin{center}
    \leavevmode
\epsfxsize=6.0cm  
\epsfbox[134 380 418 720]{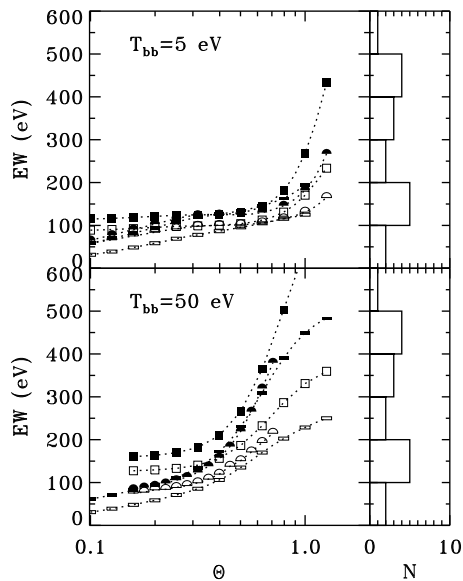} 
  \end{center}
  \caption{\em Equivalent width of the iron K-line at 6.4 keV vs. 
electron temperature, $\Theta$. Same notations as in Figure~3. 
Notice that for low temperature, the equivalent width is almost constant. 
It increases when the temperature increases due to the anisotropy effects (see
Fig.~2). For higher disc 
temperature, $\Tbb$, the anisotropy effects become important at
smaller electron temperature. Right panels
show the distribution of the equivalent widths of 
Seyfert 1s observed by ASCA (Nandra et al. 1996, Table~6). } 
 \label{fig:ew} 
\end{figure}

When the anisotropy break shifts to the spectral region around the fluorescent 
iron line at 6.4 keV, the continuum flux in the direction 
of the cold disc exceeds the flux in the observer's direction, 
causing an increase of the  equivalent width of the K-line in
comparison with the isotropic case. This effect is demonstrated 
in Figure~\ref{fig:ew} where the iron line equivalent width is plotted 
against the electron temperature for different geometries. We 
also show the distribution of the equivalent widths observed by 
{\em ASCA} (\cite{n96}).  

\begin{figure}[htbp] 
  \begin{center}
    \leavevmode
\epsfxsize=6.0cm
\epsfbox[84 380 468 720]{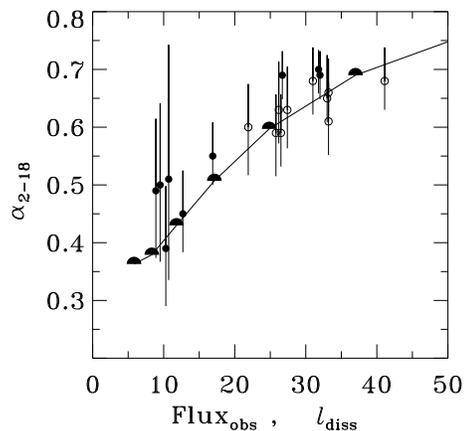}
  \end{center}
  \caption{\em Correlation between spectral index, $\a218$, and the 
continuum flux in 2 - 10 keV energy band corrected for absorption 
(in units of 10$^{-11}\ergs\sec^{-1}\cm^{-2}$)  observed
by Ginga in NGC 4151.  Filled spheres with the error bars show the  data 
from Yaqoob $\&$ Warwick (1991) 
and open spheres are the data from Yaqoob et al. (1993). 
Hemispheres and solid curve show the correlation between 
spectral index and dissipation compactness (arbitrary compactness units)
for the hemisphere-corona model. Similar correlation is 
expected also in non-thermal pair models (see Yaqoob 1992). } 
 \label{fig:correl}
\end{figure}

\section{FLUX - SPECTRAL INDEX CORRELATION} 

Some Seyfert galaxies show correlated variations in 
flux and spectral index (see, e.g., \cite{mat90}, \cite{tre90}, 
\cite{yw91}, \cite{y93}). This correlation
has been explained in the context of non-thermal pair model 
(e.g. \cite{y92}). 
Our thermal pair model shows strong correlation between compactness 
parameter and the spectral index (see Fig.~\ref{fig:al218}). 
Assuming that the size of the X/$\gamma$ source is constant there 
is a direct correspondence between $\ldiss$ and the observed flux
(here we neglect the effect of spectral changes with compactness). 
Figure~\ref{fig:correl}  shows the flux -- spectral index correlation 
observed in NGC 4151 and the  compactness -- spectral index correlation 
for a hemisphere-corona ($\ldiss$ is in arbitrary units and 
is rescaled to fit observations). It can be seen in Figure~\ref{fig:al218}, 
that for $\Tbb$=5 eV  spectral index changes dramatically from 
1.0 to 0.5 when $\ldiss$ changes by an order of magnitude from 100 to 10.

\section{SPECTRAL FITTING} 

\begin{figure}[htbp] 
  \begin{center}
    \leavevmode
\epsfxsize=6.0cm  
\epsfbox[84 380 488 650]{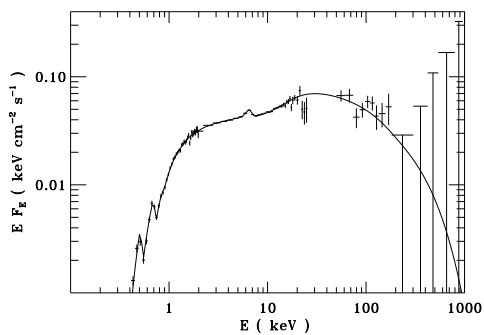}
  \end{center}
  \caption{\em The broad-band spectrum of Seyfert 1 galaxy, IC4329A. 
ROSAT, Ginga, and OSSE observations are described in Madejski et al. (1995).
The upper limits are $2\sigma$. 
Solid curve shows the best fit hemisphere-corona model spectrum. } 
 \label{fig:ic} 
\end{figure}

We implemented the ISM code into XSPEC, and used it to 
fit the broad band spectrum of the second brightest Seyfert 1 galaxy, IC4329A.  
As an example, we fitted data with theoretical spectra
from a hemisphere-corona. 
The  spectrum is absorbed by
a combination of neutral and ionized absorbers. 
It is  presently difficult  to constrain 
the temperature of the black-body radiation, $\Tbb$, 
and hence we fixed it at 10 eV. 
The best fit model parameters are  $\Te$=100$^{+10}_{-30}$ keV, and 
$\taut$=1.3$^{+0.4}_{-0.2}$. 
The quality of the data does not allow  us to  constrain the geometry 
of the emitting region. 
The data and the model spectrum are shown in Figure~\ref{fig:ic}. 

\section{CONCLUSIONS}

Several diagnostics from physical modelling of observed spectra 
are possible. 
(1) As anisotropic effects are very important, i.e. as
spectral shapes depend strongly on viewing angle, it will be possible
to set constraints on the viewing angle within the framework of 
the two-phase pair corona model, once high quality spectra become available. 
(2) The spectra also depend on the geometry of the coronal regions, so
observed  high quality  spectra can be used as diagnostics of the geometry.
Presently, it seems that active regions are favored over homogeneous
slab coronae. 
(3) Observations of  flux-index  correlations can provide constraints on the 
source compactness. 
(4) The equivalent width of the iron line can be used as a diagnostic of the 
temperature of the reprocessed (black-body) radiation which then constrains 
the size.

\section*{ACKNOWLEDGMENTS}

This research was  supported by
grants and a postdoctoral fellowship (J.P.)
from the Swedish Natural Science Research Council.

\end{document}